# Interfacial crystal Hall effect reversible by ferroelectric polarization

Ding-Fu Shao,[1,*] Jun Ding,[2] Gautam Gurung,[1] Shu-Hui Zhang,[3] and Evgeny Y. Tsymbal[1,†]

[1] *Department of Physics and Astronomy & Nebraska Center for Materials and Nanoscience,*
*University of Nebraska, Lincoln, Nebraska 68588-0299, USA*

[2] *College of Science, Henan University of Engineering, Zhengzhou 451191, People's Republic of China*

[3] *College of Mathematics and Physics, Beijing University of Chemical Technology,*
*Beijing 100029, People's Republic of China*

The control of spin-dependent properties by voltage, not involving magnetization switching, has significant advantages for low-power spintronics. Here, we predict that the interfacial crystal Hall effect (ICHE) can serve for this purpose. We show that the ICHE can occur in heterostructures composed of compensated antiferromagnetic metals and non-magnetic insulators due to reduced symmetry at the interface, and it can be made reversible if the antiferromagnet is layered symmetrically between two identical ferroelectric layers. We explicitly demonstrate this phenomenon using density functional theory calculations for three material systems: $MnBi_2Te_4/GeI_2$ and topological $In_2Te_3/MnBi_2Te_4/In_2Te_3$ van der Waals heterostructures, and $GeTe/Ru_2MnGe/GeTe$ heterostructure composed of three-dimensional materials. We show that all three systems reveal a sizable ICHE, while the latter two exhibit a quantum ICHE and ICHE, respectively, reversible with ferroelectric polarization. Our proposal opens an alternative direction for voltage controlled spintronics and offers not yet explored possibilities for functional devices by heterostructure design.

## I. INTRODUCTION

Spintronics is a research field that exploits the spin-dependent transport properties of material structures in electronic devices, such as non-volatile memories and logics [1]. Spontaneous magnetization $M$ is often employed as a state variable in these devices, and its orientation is used for distinguishing non-volatile states. "Standard" methods of controlling magnetization, such as a magnetic field or a spin torque, require substantial electric currents and thus suffer from large energy dissipation. An electric control of magnetization by voltage has significant advantages for low-power spintronics [2]. However, the electric field does not break time-reversal symmetry $\hat{T}$ and hence on its own is not sufficient to reverse the magnetization.

An electric field breaks the space-inversion symmetry $\hat{P}$ and leads to electric polarization $P$ in insulators. In multiferroic insulators, where the spontaneous polarization coexists with a magnetic order, a non-volatile electric control of magnetization is possible due to the magnetoelectric coupling [3,4]. However, the applicable room-temperature multiferroics are rare. Another approach is to exploit the interfacial magnetoelectric coupling in heterostructures of ferroelectric and ferromagnetic materials [5], where reversal of ferroelectric polarization affects the interface magnetization through the electrostatic doping effect. However, this approach requires precise engineering of stoichiometry [6-9] thickness [10,11], or a non-collinear magnetic order [12], which may be complicated for the design of realistic devices.

For low-power memory and logic devices, it would be beneficial to realize an electric field control of the spin-dependent properties without magnetization switching. For example, it has been theoretically proposed [13] and experimentally demonstrated [14] that the spin polarization in a nonmagnetic two-dimensional (2D) material can be induced and tuned by gate voltage due to proximity of a ferromagnet. However, while this approach may potentially offer a low-power spintronic device, it relies on a *volatile* control of the spin polarization which appears only in the presence of gate voltage.

In contrast, using spin-textured ferroelectrics [15] allows control of spin-dependent properties by voltage in a *non-volatile* way due to a switchable spontaneous polarization of the ferroelectric. In certain ferroelectric materials, such as GeTe [16,17], the spin texture is coupled to ferroelectric polarization and reversed with polarization switching. This property offers a possibility of a bi-stable voltage control of the spin-dependent properties, such as anomalous Hall effect (AHE) [18,19], in an appropriate magnetic conducting system.

The intrinsic AHE is driven by the Berry curvature, $\mathbf{\Omega}$, an intrinsic property of a material arising from its spin-dependent band structure [20-22]. The anomalous Hall conductance (AHC) is determined by the integral of $\mathbf{\Omega}$ weighted with the Fermi distribution function over the whole Brillouin zone. It is non-zero for a system with no symmetry operation $\hat{O}$, with respect to which the Berry curvature is antisymmetric, i.e. $\hat{O}\mathbf{\Omega}(\mathbf{k}) = -\mathbf{\Omega}(\hat{O}\mathbf{k})$. The AHC changes sign with magnetization reversal, which is equivalent to the application of time-reversal symmetry operation $\hat{T}$ to the system, and hence transforms the Berry curvature as $\hat{T}\mathbf{\Omega}(\mathbf{k}) = -\mathbf{\Omega}(-\mathbf{k})$. The sign of the Berry curvature can be also changed by other symmetry operations which do not reorient the magnetic moments. This property has recently been discussed in conjunction with collinear antiferromagnets such as $RuO_2$ [23, 24], $NiF_2$ [25], $CoNb_3S_6$ [26], and ultra-thin $SrRuO_3$ [27], where the AHE is supported by the non-centrosymmetric crystal structure. This phenomenon was coined the crystal Hall effect (CHE) [23].

An interesting consequence of the CHE is that the switching of the positions of non-magnetic atoms is equivalent to the application of the symmetry operation that changes sign of the Berry curvature, and hence reverses the AHC. The reversal of the AHC does not require magnetic moment switching typical for the conventional AHE. Such a functional property could be useful for applications in low-power spintronics, eliminating the need for large energy-dissipating electric currents to switch the AFM order parameter [28].

Unfortunately, there are no means to exploit this property in bulk materials, due to no external stimulus which could possibly switch the non-magnetic sublattice in a metal alone. This property however can be obtained in a heterostructured material where the magnetic group symmetry is affected across the interface due to the proximity effect. This brings a practical perspective to realize a reversible CHE.

In this work, we demonstrate that the CHE may occur not only in bulk materials but also in heterostructures composed of compensated AFM metals and non-magnetic insulators due to reduced symmetry at the interface. Different from the bulk CHE, such an interfacial CHE (ICHE) does not require the non-symmetric atomic positions in the bulk antiferromagnet. The interfacial proximity effect alone breaks the antisymmetry of the Berry curvature and produces an interfacial crystal Hall conductance (ICHC). We further show that using ferroelectric materials for non-magnetic insulators in the heterostructure allows the realization of the reversible ICHE, where the ICHC changes sign with ferroelectric polarization switching. We explicitly demonstrate these phenomena using first-principles density functional theory (DFT) calculations (see Appendix A for details) for three material systems: a MnBi$_2$Te$_4$/GeI$_2$ van der Waals heterostructure, where we show the emergence of a sizable ICHC; an In$_2$Te$_3$/MnBi$_2$Te$_4$/In$_2$Te$_3$ topological van der Waals heterostructure, where we predict a quantized ICHC reversible by ferroelectric polarization; and a GeTe/Ru$_2$MnGe/GeTe layered heterostructure composed of three-dimensional (3D) materials, where we predict a reversible ICHC. Our prediction offers a feasible solution for the non-volatile electric switching of a spin-dependent transport property, and hence opens an alternative direction in voltage-controlled spintronics.

## II. RESULTS

### A. General considerations

Compensated antiferromagnets are magnets exhibiting symmetry $\hat{O}$ that prevents net magnetization. This symmetry may be a combination of time reversal $\hat{T}$ and a crystal symmetry operation or a combination of multiple crystal symmetry operations, which enforces the antisymmetry of the

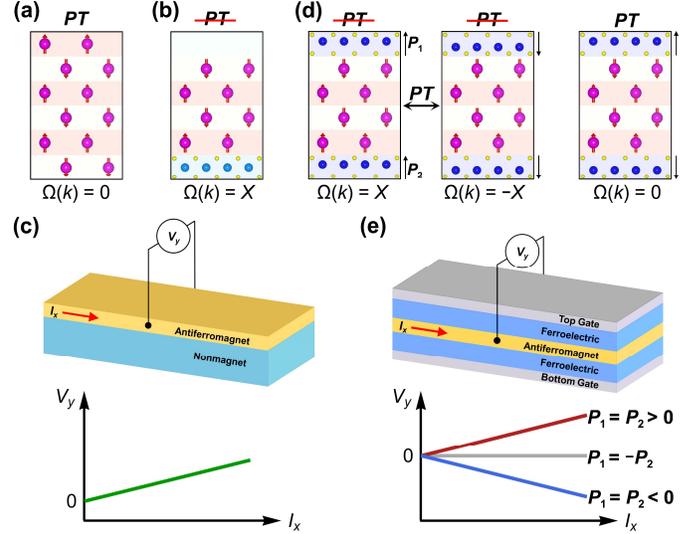

FIG. 1: **(a)** A schematic of A-type antiferromagnet with preserved $\hat{P}\hat{T}$ symmetry, enforcing $\mathbf{\Omega}(\mathbf{k}) = 0$. **(b)** A schematic of a heterostructure where the AFM layer in (a) is placed on a non-magnetic substrate. The interface breaks the $\hat{P}\hat{T}$ symmetry resulting in non-zero $\mathbf{\Omega}(\mathbf{k})$ and hence ICHE. **(c)** A schematic of a device for observing the ICHE. Longitudinal in-plane charge current $I_x$ is passing through the AFM layer which generates the transverse Hall voltage $V_y$ with an I-V characteristic shown at the bottom. **(d)** A schematic of a ferroelectric/antiferromagnet/ferroelectric heterostructure where the AFM layer in (a) is inserted between two identical ferroelectric layers. If the polarizations of the top and bottom layers ($\mathbf{P}_1$ and $\mathbf{P}_2$) are parallel, the $\hat{P}\hat{T}$ symmetry is broken resulting in non-zero $\mathbf{\Omega}(\mathbf{k})$ and hence ICHE (left). Switching $\mathbf{P}_1$ and $\mathbf{P}_2$ simultaneously equivalent to applying $\hat{P}\hat{T}$ symmetry operation which changes sign of $\mathbf{\Omega}(\mathbf{k})$ (center). If $\mathbf{P}_1$ and $\mathbf{P}_2$ are antiparallel, the $\hat{P}\hat{T}$ symmetry is preserved enforcing $\mathbf{\Omega}(\mathbf{k}) = 0$ (right). **(e)** A schematic of a device for observing the reversible ICHE with I-V characteristics shown at the bottom. The top and bottom gates are used to control $\mathbf{P}_1$ and $\mathbf{P}_2$ separately.

Berry curvature $\mathbf{\Omega}$ and thus prohibits the AHE in a bulk antiferromagnet.

However, the above-mentioned symmetry $\hat{O}$ is expected broken at interface and thus all compensated antiferromagnets are to exhibit an ICHE in appropriate heterostructures. As a representative example, we focus on A-type antiferromagnets with an out-of-plane AFM order parameter, where the two sublattices are connected by $\hat{P}\hat{T}$ symmetry (Fig. 1(a)). This type of the AFM order has been found in many compounds, such as 3D metals Ru$_2$MnGe [29, 30], MnPd$_2$ [31, 32], and CaCo$_2$As$_2$ [33], and 2D van der Waals semiconductors MnBi$_2$Te$_4$ [34] and CrI$_3$ [35]. In these materials, the Berry curvature is zero everywhere in the Brillouin zone enforced by $\hat{P}\hat{T}\mathbf{\Omega}(\mathbf{k}) = -\mathbf{\Omega}(\mathbf{k})$, and hence the CHE is prohibited. However, in a heterostructure composed of such an antiferromagnet and a non-magnet (Fig. 1(b)), the $\hat{P}\hat{T}$ symmetry is broken at the interface, resulting in a non-zero $\mathbf{\Omega}(\mathbf{k})$ and a non-vanishing ICHE (Fig. 1(c)).

By inserting a slab of such an antiferromagnet between two identical ferroelectric layers with out-of-plane polarizations (Fig. 1(d)), we design a heterostructure for a reversible ICHE. The polarization of the top and bottom layers ($\mathbf{P}_1$ and $\mathbf{P}_2$) can be controlled by the top and bottom gates separately in a device schematically shown in Figure 1(e). These polarizations control the magnetic group symmetry of the heterostructure that determines the ICHE. When $\mathbf{P}_1$ and $\mathbf{P}_2$ are parallel, the $\hat{P}\hat{T}$ symmetry is broken and hence $\mathbf{\Omega}(\mathbf{k})$ is non-zero resulting in the ICHE. Switching $\mathbf{P}_1$ and $\mathbf{P}_2$ simultaneously is equivalent to applying the $\hat{P}\hat{T}$ symmetry operation to the heterostructure, which changes sign of $\mathbf{\Omega}(\mathbf{k})$ and reverses the ICHC. When $\mathbf{P}_1$ and $\mathbf{P}_2$ are antiparallel, the $\hat{P}\hat{T}$ symmetry is preserved, leading to zero Hall voltage. Therefore, the three non-volatile ICHE states coupled to ferroelectric polarization are realized in the spintronic device based on this heterostructure.

## B. ICHE in a MnBi$_2$Te$_4$/GeI$_2$ bilayer

To demonstrate these properties, we first consider a 2D van der Waals heterostructure composed of four-monolayer AFM MnBi$_2$Te$_4$ deposited on non-magnetic monolayer GeI$_2$ (Fig. 2(a)). Bulk MnBi$_2$Te$_4$ is an A-type AFM topological insulator with out-of-plane magnetic moments. It preserves the $\hat{P}\hat{T}$ symmetry and has the Néel temperature ($T_N$) of 25 K [34]. Bulk GeI$_2$ is a centrosymmetric wide gap semiconductor (band gap is ~2.5 eV) [36, 37] with the cleavage energy lower than that in graphite [38].

Both freestanding MnBi$_2$Te$_4$ and GeI$_2$ have the symmetry operations preventing linear or nonlinear AHE [39, 40]. However, these symmetries are broken in the MnBi$_2$Te$_4$/GeI$_2$ heterostructure. Figure 2(b) shows the calculated band structure of MnBi$_2$Te$_4$/GeI$_2$, where the bands near the Fermi energy ($E_F$) originate from MnBi$_2$Te$_4$ (see also Fig. 5(b) in Appendix B). The Kramers degeneracy enforced by $\hat{P}\hat{T}$ symmetry is lifted in this heterostructure, as seen from the small but non-negligible band splitting in Fig. 2(b). We note that the change in the crystal structure of MnBi$_2$Te$_4$ is vanishingly small, and Mn magnetic moments are the same as these in a freestanding MnBi$_2$Te$_4$ slab, indicating that the reduced symmetry is purely due to the interfacial proximity effect. This small band splitting is sufficient to produce a sizable Berry curvature $\mathbf{\Omega}(\mathbf{k}) = \sum_n f_{n\mathbf{k}}\mathbf{\Omega}_{n\mathbf{k}}$, where $f_{n\mathbf{k}}$ in the Fermi distribution function and $\mathbf{\Omega}_{n\mathbf{k}}$ is the Berry curvature of the $n$-th band given by [20,22]:

$$\mathbf{\Omega}_{n\mathbf{k}} = -2\mathrm{Im}\sum_{m \neq n}\frac{\left\langle n\left|\frac{\partial \hat{H}}{\partial k_x}\right|m\right\rangle\left\langle m\left|\frac{\partial \hat{H}}{\partial k_y}\right|n\right\rangle}{(E_{n\mathbf{k}} - E_{m\mathbf{k}})^2}.$$

Here $\hat{H}$ is the Hamiltonian of the system and $E_{n\mathbf{k}}$ is the $n$-th band energy at wave vector $\mathbf{k}$. Figure 2(c) shows the calculated $\mathbf{k}$-dependent Berry curvature $\mathbf{\Omega}(\mathbf{k})$ at energy $E_1$ located 64 meV below the top of the valence band (Fig 2(b)). The sizable $\mathbf{\Omega}(\mathbf{k})$ seen near the Brillouin zone center is not antisymmetric. This provides a non-vanishing ICHC $\sigma_{xy}$ given by [19]

$$\sigma_{xy} = -\frac{e^2}{h}\int_{BZ}\frac{d^2\mathbf{k}}{2\pi}\mathbf{\Omega}(\mathbf{k}).$$

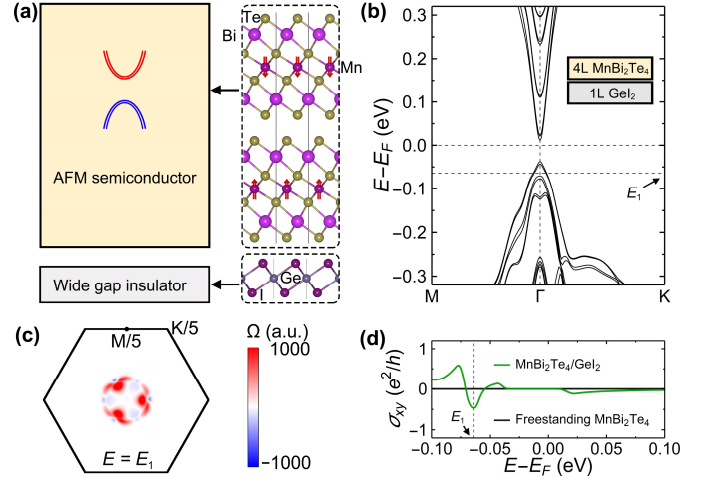

**FIG. 2: (a)** A schematic of the MnBi$_2$Te$_4$/GeI$_2$ van der Waals heterostructure. Red arrows denote magnetic moments of Mn atoms. **(b)** Band structure of the MnBi$_2$Te$_4$/GeI$_2$ heterostructure. **(c)** Berry curvature in atomic units (a.u.) at the center of the 2D Brillouin zone at energy $E_1$ indicated in (b). **(d)** The calculated ICHC as a function of energy near $E_F$.

Figure 2(d) shows that $\sigma_{xy}$ is zero at $E_F$, indicating a trivial insulating state of this heterostructure. In experiment, however, the $E_F$ of a few-layer MnBi$_2$Te$_4$ is usually located slightly below the valence band maximum [41]. As seen from Fig. 2(d), $\sigma_{xy}$ is non-zero for $E_F$ shifted to the valence band. This is in contrast to a freestanding MnBi$_2$Te$_4$ (Fig. 5(a)), where $\sigma_{xy}$ is zero for all energies due to $\hat{P}\hat{T}$ symmetry (Fig. 2(d)). Thus, the presence of GeI$_2$ breaks the $\hat{P}\hat{T}$ symmetry and produces an ICHE.

Experimentally non-zero $\sigma_{xy}$ has been observed in four- and six-layer MnBi$_2$Te$_4$ on substrates [41, 42], which may serve as the evidence of the ICHE. The ICHC can be enhanced to ~0.5 $e^2/h$ at $E_1$ (Fig. 2d) by adjusting the $E_F$ of MnBi$_2$Te$_4$ with a gate voltage similar to that done in Ref. 41.

## C. Quantized ICHE in a GeI$_2$/In$_2$Te$_3$/MnBi$_2$Te$_4$/In$_2$Te$_3$/GeI$_2$ heterostructure

An ICHE heterostructure can be engineered to make the ICHC quantized. A quantum AHE has been predicted at zero magnetic field in MnBi$_2$Te$_4$ with AFM layer ordering [43-47] but the experimental realizations are missing. Here, we show the emergence of a quantum ICHE in a designed heterostructure, where MnBi$_2$Te$_4$ is sandwiched between two identical ferroelectric layers. For the ferroelectric component, we choose In₂Te₃ due to its lattice matching to MnBi₂Te₄ (Appendix B). In₂Te₃ belongs to the group of 2D materials In₂X₃ (X = S, Se, Te), where ferroelectricity with out-of-plane polarization has been predicted [48] and in the case of In₂Se₃, experimentally confirmed [49, 50].

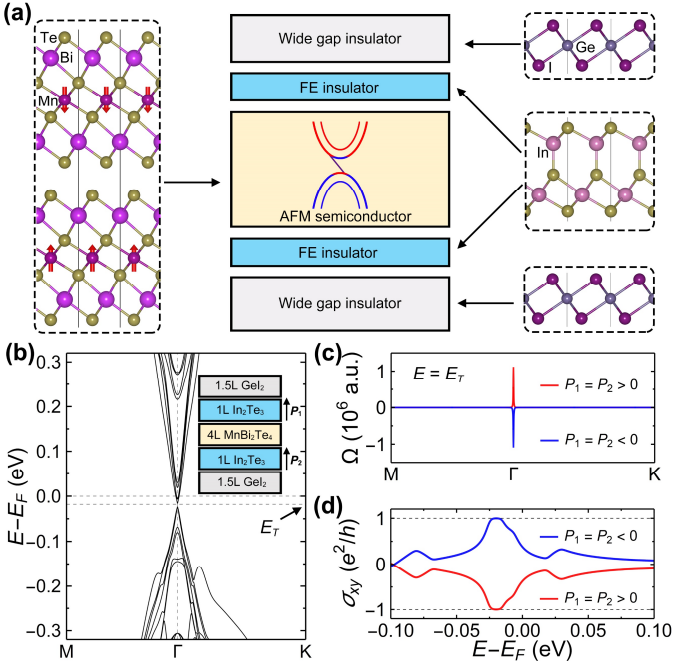

**FIG. 3:** **(a)** A schematic of the GeI₂/In₂Te₃/MnBi₂Te₄/In₂Te₃/GeI₂ van der Waals topological heterostructure for reversible quantized ICHC. **(b)** The band structure of this heterostructure with four-monolayer MnBi₂Te₄ for polarizations of top and bottom In₂Te₃ monolayers are parallel. **(c)** The Berry curvature in atomic units (a.u.) of this heterostructure along the high symmetric directions at energy $E_T$ denoted in **(b)**. **(d)** The ICHC as a function of energy for different polarization states.

Specific calculations are performed for a GeI₂/In₂Te₃/MnBi₂Te₄/In₂Te₃/GeI₂ heterostructure (Figs. 3(a) and 5(d)), where a GeI₂ capping layer is used to eliminate a band overlap with vacuum (see Appendix B for details). This is reminiscent to experiments where capping layers are used to prevent open surfaces in van der Waals heterostructures. For parallel polarization of the top and bottom In₂Te₃ layers ($P_1$ and $P_2$), we find that polarization charges generate a built-in electric field resulting in a larger band splitting and band bending, reducing the band gap (Figs. 3(b) and 5(d)). In the heterostructure with four-layer MnBi₂Te₄, a small band gap of ~9 meV is obtained. A huge Berry curvature is calculated within this gap at the Γ point (Fig. 3(c)). Simultaneous switching of $P_1$ and $P_2$ changes sign of the Berry curvature, due to this switching being equivalent to applying the $\hat{P}\hat{T}$ symmetry operation to the heterostructure. The calculated Chern numbers are ±1 depending on the polarization direction of the In₂Te₃ layers, indicating the topological state of this heterostructure. A quantized ICHC reversible with ferroelectric polarization is found within the topological band gap, as shown in Figure 3(d). With the sizable Néel temperature and vanishing net magnetization, the quantized ICHC in this heterostructure is expected to be robust against thermal fluctuations and magnetic perturbations, and hence may be observed at relatively high temperature.

## D. ICHE in a GeTe/Ru₂MnGe/GeTe system

A broad range of high temperature 3D antiferromagnets and ferroelectrics allow the design of heterostructures with the ICHE applicable in realistic spintronic devices. Here, we explore a GeTe/Ru₂MnGe/GeTe system (Fig. 4(a)), where Ru₂MnGe is a 3D Heusler alloy with $T_N$ of 316 K and out-of-plane A-type AFM order within the (111) planes (Fig. 6) [29, 30]. GeTe is a ferroelectric semiconductor with the Curie temperature ($T_C$) of about 700 K and spontaneous polarization arising from the polar displacement of Ge atoms with respect to Te atoms (Fig. 7) [16, 17, 51]. This heterostructure is composed of six-layer Ru₂MnGe and six-layer GeTe slabs and has symmetric Ru-Te interfacial terminations at the top and bottom interfaces.

Strong bonding across the interfaces in this heterostructure produces atomic displacements in the AFM layer, which enhances symmetry breaking when the out-of-plane polarization of the top and bottom GeTe layers ($P_1$ and $P_2$) are parallel. This leads to small changes the magnetic moment of the interfacial Mn atoms. We find that the Mn magnetic moments in the center of the Ru₂MnGe layer are close to the bulk moments of 3.8 μB/Mn, while the Mn magnetic moments near the interfaces are slightly influenced by the polarization of GeTe, due to the interfacial structural relaxation and accumulated screening charges induced by the polarization (Fig. 4(b)). Specifically, the Mn moments near an interface are slightly enhanced when the polarization points toward this interface, and are slightly reduced when the polarization points away from this interface. The difference between the Mn moments in the heterostructure and these in the bulk is small (< 0.1 μB/Mn). When the polarizations $P_1$ and $P_2$ are parallel, there is a small net magnetization of the Ru₂MnGe layer ~0.026 μB due to broken $\hat{P}\hat{T}$ symmetry. The presence of the small magnetization is reflected in the calculated spin-dependent DOS shown in Fig. 4(c). Switching $P_1$ and $P_2$ simultaneously is equivalent to a $\hat{P}\hat{T}$ symmetry operation, which interchanges the magnetic moments from top to bottom within the Ru₂MnGe layer (Fig. 4(b)) and swaps the DOS contributed by up- and down-spins (Figs. 4(c)).

Figure 4(d) shows the band structure of GeTe/Ru₂MnGe/GeTe for parallel $P_1$ and $P_2$. There are several bands crossing the Fermi energy, which are majorly contributed by the Ru-$5d$ electrons (Fig. 9(c)). Notable is the band splitting enforced by the broken $\hat{P}\hat{T}$ symmetry, which supports a non-vanishing Berry curvature (Fig. 4(e)) and sizable ICHC reversible by ferroelectric polarization (Fig. 4(f)). For $P_1 = P_2 > 0$, $\sigma_{xy} = -1.4\ e^2/h$ at $E_F$, which is comparable to AHC of 2D ferromagnetic metal Fe$_3$GeTe$_2$ [52, 53]. The ICHC can be enhanced to $\sigma_{xy} = -5.5\ e^2/h$ by proper electron doping. When $P_1$ and $P_2$ are reversed the ICHC changes sign. For $P_1$ and $P_2$ having opposite directions, the $\hat{P}\hat{T}$ symmetry enforces Kramers degeneracy and vanishing ICHC (Fig. 4(f)). We note that different interfacial configurations do not affect our conclusions about the reversible ICHE, as long as the symmetry of the designed heterostructure is guaranteed (see Appendix D and Fig. 8 and for details).

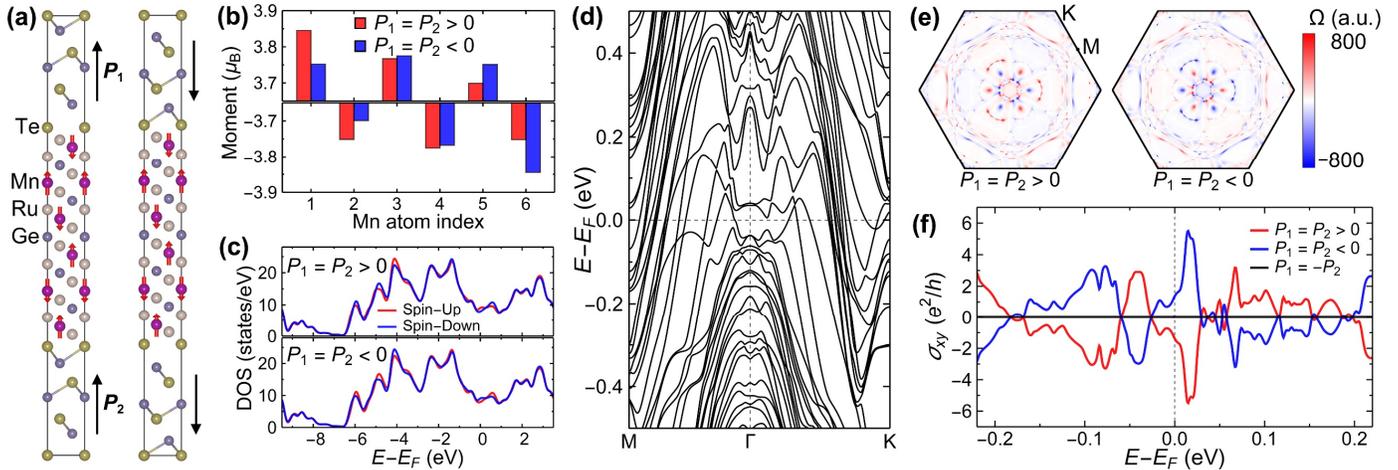

**FIG. 4:** **(a)** The atomic structure of GeTe/Ru$_2$MnGe/GeTe heterostructure for parallel out-of-plane polarizations of the top and bottom GeTe layers ($P_1$ and $P_2$), for $P_1$ and $P_2$ pointing up (left) and down (right). **(b)** The calculated magnetic moments for each Mn atoms for $P_1$ and $P_2$ pointing up ($P_1 = P_2 > 0$) and down ($P_1 = P_2 < 0$). Here Mn atoms in the Ru$_2$MnGe slab **(a)** are indexed from 1 (bottom) to 6 (top). **(c)** The spin-dependent density of states of the heterostructure for $P_1$ and $P_2$ pointing up ($P_1 = P_2 > 0$, top panel) and down ($P_1 = P_2 < 0$, bottom panel) calculated without spin-orbit coupling. **(d)** The band structure of this heterostructure in the presence of spin-orbit coupling. **(e)** The Berry curvature in atomic units (a.u.) at $E_F$ in the 2D Brillouin zone for $P_1$ and $P_2$ pointing up ($P_1 = P_2 > 0$, left panel) and down ($P_1 = P_2 < 0$, right panel). **(f)** The ICHC as a function of energy for different polarization states.

## III. DISCUSSION

Most AFM compounds are compensated antiferromagnets, where the AHE is forbidden by symmetry of bulk materials, making believe that these materials are not useful for spintronic applications due to no means to sense their order parameter in nanoscale devices. On the contrary, our predictions demonstrate that these compensated antiferromagnets, if properly interfaced, can be become practical for these applications due to non-vanishing ICHE. This prediction significantly broadens the range of antiferromagnets that can be employed in spintronics.

Based on the available material choice, the heterostructures supporting the reversable ICHE can have different types. As discussed in Appendix E and shown in Fig. 10, in the heterostructures, the ferroelectric layers can have in-plane polarization and the AFM layers can have various types of magnetic ordering. As long as the designed heterostructure does not have the symmetry operation to prevent the ICHE, and the ferroelectric switching in this heterostructure is equivalent to a symmetry operation which changes sign of the Berry curvature, the ferroelectric-reversible ICHE is supported.

The predicted reversibility of the ICHE by electric fields offers a solution of the critical problem of spintronics: Non-volatile electric switching of a spin-dependent transport property by voltage. Our prediction shows that the ICHE, a spin-dependent transport property, can be reversed with ferroelectric polarization in a designed heterostructure. This functionality cannot be realized in the systems considered previously, where the AHE can only be reversed through magnetization switching.

At the end, we would like to mention that the proposed heterostructures are feasible for the practical realization. All the components in these heterostructures have been successfully fabricated through exfoliation or epitaxial growth [17, 41, 42, 49, 54]. These methods can also be used to synthesize the required multicomponent layered systems.

## IV. CONCLUSIONS

In conclusion, we have predicted the emergence of the interfacial crystal Hall effect in heterostructures composed of compensated antiferromagnetic and non-magnetic metals. The effect occurs due to the broken symmetry between the two AFM sublattices resulting from the proximity effect. This result implies that the AHE can be observed in a broad range of properly interfaced compensated antiferromagnets, which have previously been considered as incompatible with the AHE. We have also proposed the ICHE reversible with ferroelectric polarization in designed AFM heterostructures with symmetric top and bottom ferroelectric layers. The predicted phenomena have been demonstrated for the specific material systems. For MnBi$_2$Te$_4$/GeI$_2$, we confirmed a sizable ICHE. For In$_2$Te$_3$/MnBi$_2$Te$_4$/In$_2$Te$_3$ and GeTe/Ru$_2$MnGe/GeTe, we predicted a quantized ICHC for the former and a sizable ICHC for the latter, both reversible by ferroelectric polarization. This prediction offers a feasible solution for the non-volatile electric switching of the spin-dependent transport property, and hence opens an alternative direction for voltage controlled spintronics. We hope therefor that our theoretical predictions will motivate experimentalists to explore the predicted ICHE.

## ACKNOWLEDGMENTS

The authors thank Bo Li for helpful discussions. This work was supported by the National Science Foundation (NSF) through the Nebraska MRSEC program (grant DMR-1420645) and the DMREF program (grant DMR-1629270). Computations were performed at the University of Nebraska Holland Computing Center.

## APPENDIX A: COMPUTATIONAL METHODS

First-principles calculations are performed with the projector augmented-wave (PAW) method [55] implemented in the VASP code [56]. The exchange and correlation effects are treated within the generalized gradient approximation (GGA) [57]. We use the plane-wave cut-off energy of 350 eV and a 16

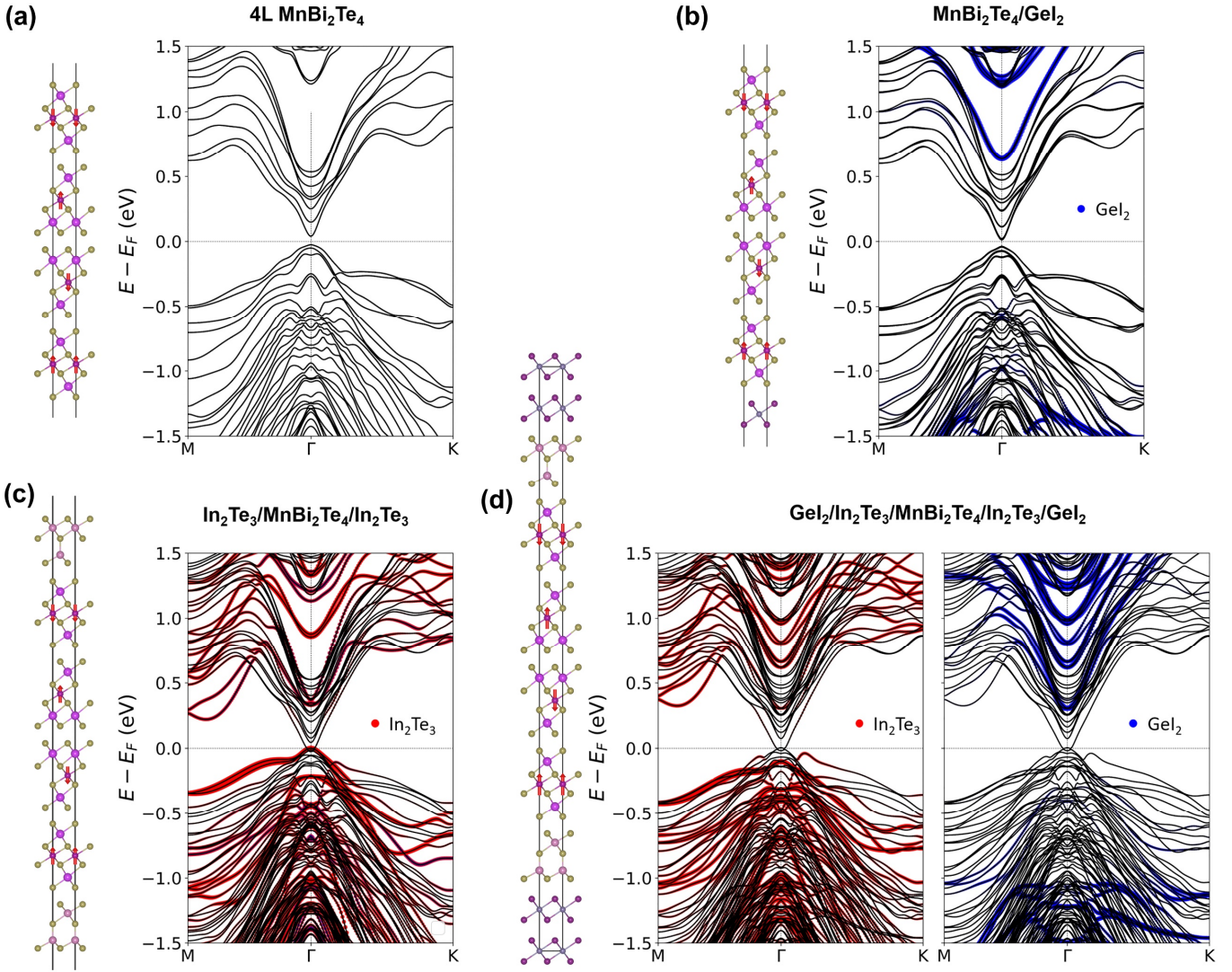

FIG. 5: (**a**) The unit cell (left) and band structure of a four-layer MnBi$_2$Te$_4$ slab. (**b**) The unit cell (left) and band structure of MnBi$_2$Te$_4$/GeI$_2$ heterostructure with four MnBi$_2$Te$_4$ layers. (**c**) The unit cell (left) and band structure of In$_2$Te$_3$/MnBi$_2$Te$_4$/In$_2$Te$_3$ heterostructure with four MnBi$_2$Te$_4$ layers. (**d**) The unit cell (left) and band structure of the GeI$_2$/In$_2$Te$_3$/MnBi$_2$Te$_4$/In$_2$Te$_3$/GeI$_2$ heterostructure.

**Table 1:** Calculated magnetic anisotropy of the ICHE systems.

|  | $E_z - E_x$ (meV/cell) |
|---|---|
| 4L MnBi$_2$Te$_4$ | -1.8 |
| MnBi$_2$Te$_4$/GeI$_2$ | -1.6 |
| GeI$_2$/In$_2$Te$_3$/MnBi$_2$Te$_4$/In$_2$Te$_3$/GeI$_2$ | -1.5 |
| GeTe/Ru$_2$MnGe/GeTe | -7.5 |

× 16 × 1 $k$-point mesh in the irreducible Brillouin zone. The GGA+U functional [58,59] with $U_{eff}$ = 3 eV on Mn $3d$ orbitals is included in all calculations. The van der Waals corrections as parameterized within the semiempirical DFT-D3 method [60] are used in the calculations of the 2D van der Waals heterostructures. The rhombohedral stacking is considered in all heterostructures. The in-plane lattice constant of MnBi$_2$Te$_4$-based heterostructures is constrained to the calculated in-plane lattice constant of MnBi$_2$Te$_4$, while the in-plane lattice constant of the GeTe/Ru$_2$MnGe/GeTe heterostructure is constrained to the calculated in-plane lattice constant of GeTe. The out-of-plane lattice parameters and atomic positions are relaxed until the force on each atom is less than 0.001 eV/Å. The electronic structure of the GeI$_2$/In$_2$Te$_3$/MnBi$_2$Te$_4$/In$_2$Te$_3$/GeI$_2$ system is further corrected by the modified Becke-Johnson (mBJ) functional [61]. Unless mentioned in the text, the spin-orbit coupling is included in all calculations of the electronic properties.

The maximally localized Wannier functions [62] are used to obtain the tight-binding Hamiltonian within the Wannier90 code [63]. The interfacial crystal Hall conductance is calculated using the Wanniertools code [64]. A 1000 × 1000 $k$-point mesh is used to achieve the convergence of the interfacial crystal Hall conductance.

Figures are plotted using the VESTA [65], gnuplot [66], and SciDraw [67] software.

## APPENDIX B: BAND STRUCTURE OF MnBi$_2$Te$_4$-BASED SYSTEMS

The atomic structures of MnBi$_2$Te$_4$, In$_2$Te$_3$, and GeI$_2$ can be found in Figures 2(a) and 3(a). Their calculated lattice parameters are 4.353 Å, 4.392 Å, and 4.296 Å, respectively, consistent with the previous reports [34, 38, 48].

Four systems are considered: (1) four-layer MnBi$_2$Te$_4$; (2) MnBi$_2$Te$_4$/GeI$_2$ bilayer; (3) GeI$_2$/In$_2$Te$_3$/MnBi$_2$Te$_4$/In$_2$Te$_3$/GeI$_2$ layered heterostructure; and (4) GeTe/Ru$_2$MnGe/GeTe superlattice. In the calculations of systems (1)-(3), we assume the presence of a 20 Å vacuum layer separating the top and bottom surfaces. For all heterostructures, we calculate the magnetocrystalline anisotropy energy (Table 1) and find that the Néel vector is pointing along the out-of-plane direction.

Figure 5(a) shows the atomic structure and the calculated band structure of a freestanding four-layer MnBi$_2$Te$_4$. In this case, the preserved $\hat{P}\hat{T}$ symmetry enforces the double degeneracy in the band structure. The band gap is calculated to be 65 meV, leading to a trivial insulator phase.

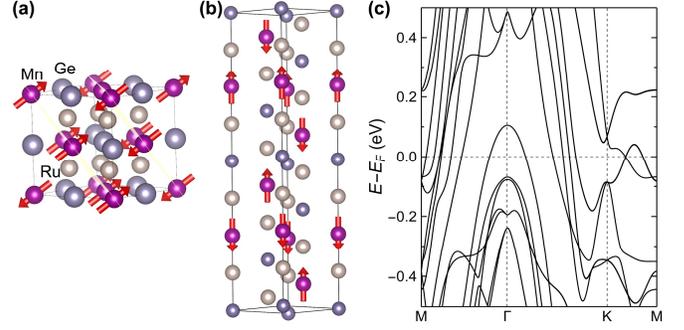

**FIG. 6**: The cubic unit cell (**a**), the hexagonal magnetic cell (**b**), and the calculated band structure (**c**) of bulk Ru$_2$MnGe

Figure 5(b) shows the atomic structure and the calculated band structure of the MnBi$_2$Te$_4$/GeI$_2$ heterostructure. It is seen that GeI$_2$ does not contribute the band structure near $E_F$. In this case, the $\hat{P}\hat{T}$ symmetry is broken and the degeneracy in the band structure is lifted, resulting in nonvanishing Berry curvature.

To calculate the band structure of In$_2$Te$_3$/MnBi$_2$Te$_4$/In$_2$Te$_3$ (Fig. 5 (c) left), the dipole correction [68] is applied to adjust the misalignment between the vacuum levels on the different sides of this heterostructure due to the intrinsic electric polarization of In$_2$Te$_3$. Figure 5(c) shows the calculated band structure for parallel polarization of the top and bottom In$_2$Te$_3$ layers. The presence of polarization breaks the $\hat{P}\hat{T}$ symmetry and produces band splitting. The polarization charge generates a built-in electric field across the heterostructure, resulting in a strong band bending which leads to an overlap of the conduction band minimum (CBM) and the valence band maximum (VBM). The VBM of In$_2$Te$_3$ is shifted above $E_F$ due to the polarization charges on the open surfaces, leading to the trivial metallic phase. This behavior has also been found for the MnBi$_2$Te$_4$/In$_2$X$_3$ bilayer heterostructure studied recently [69].

To eliminate this problem, we use three layers of GeI$_2$ instead of vacuum to separate the top and bottom surfaces in the In$_2$Te$_3$/MnBi$_2$Te$_4$/In$_2$Te$_3$ heterostructure (Fig. 5(d) left). As seen from the band structure in Figure 5(d), the VBM of In$_2$Te$_3$ is shifted down from $E_F$. This guarantees that the topological band structure of MnBi$_2$Te$_4$ near $E_F$ is unaffected by the trivial metallic phase. The band structure of the GeI$_2$/In$_2$Te$_3$/MnBi$_2$Te$_4$/In$_2$Te$_3$/GeI$_2$ heterostructure is further corrected by the modified Becke-Johnson (mBJ) functional [61] and plotted in Figure 3(b).

## APPENDIX C: CALCULATED PROPERTIES OF BULK Ru$_2$MnGe AND GeTe

Ru$_2$MnGe is a cubic Heusler alloy with magnetic moments of 3.8 μ$_B$ per Mn atom pointing along the [111] direction and

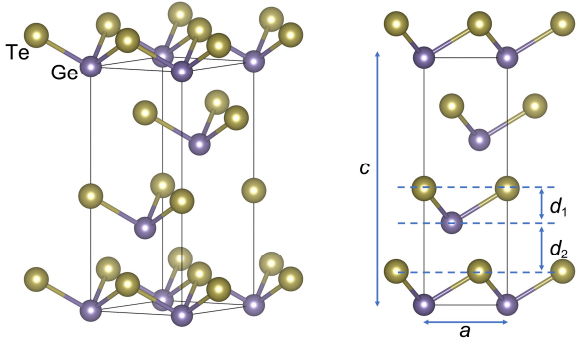

Te
Ge

$c$

$d_1$
$d_2$

$a$

**FIG. 7:** The conventional unit cell of GeTe.

lying within the (111) planes antiparallel between the successive planes (Fig. 6 (a)). This A-type antiferromagnet has the magnetic moment of 3.8 $\mu_B$ per Mn atom, the Néel temperature of 316 K, and the lattice parameter of 5.985 Å in the cubic unit cell [29, 30]. The conventional unit cell for AFM state is hexagonal containing six $Ru_2MnGe$ layers, where the out-of-plane direction is the [111] direction of the cubic cell (Fig. 6(b)). Our DFT calculations predict the magnetic moment of 3.79 $\mu_B$ per Mn atom and the lattice parameter of 6.041 Å for cubic unit cell of $Ru_2MnGe$ (corresponding to the in-plane lattice parameter of 4.272 Å for the hexagonal cell). These values are close to those found in the experiments [29, 30], indicating that our GGA+U calculation correctly reproduces the physical properties of $Ru_2MnGe$. The calculated band structures of bulk $Ru_2MnGe$ (Fig. 6 (c)) correctly reproduces the metallic ground state of $Ru_2MnGe$.

Figure 7 shows the hexagonal conventional unit cell of bulk GeTe. The ferroelectric polarization of GeTe is due to the out-of-plane polar displacement of Ge atom, which is reflected in non-zero $\Delta z_{Ge} = (d_2 - d_1)/2$, where $d_1$ and $d_2$ are the out-of-plane distance between Ge layer and the top and bottom Te layers, respectively. The in-plane ($a$) and out-of-plane ($c$) lattice constants, and the $\Delta z_{Ge}$ of bulk GeTe are calculated to be $a = 4.224$ Å, $c = 10.882$ Å, and $\Delta z_{Ge} = 0.337$ Å, which are slightly larger than these measured in experiment [51], but are consistent with previous calculations [16, 17].

## APPENDIX D: CALCULATED PROPERTIES OF GeTe/$Ru_2MnGe$/GeTe

The lattice mismatch between $Ru_2MnGe$ and GeTe is very small. In the GeTe/$Ru_2MnGe$/GeTe heterostructure, the in-plane lattice constant is constrained to 4.224 Å, the calculated in-plane lattice constant of bulk GeTe. We find that using full structural relaxation strongly suppresses the polarization of the GeTe layer (Figs. 8(b)). The similar issue has been previously reported in Refs. [70, 71]. To maintain the ferroelectric state of GeTe in the GeTe/$Ru_2MnGe$/GeTe heterostructure during the structural relaxation, we fix bulk positions of Ge and Te atoms in the interior of the GeTe layer (the atoms within the dashed box in Fig. 8(a)), and use the resulting relaxed structure for further electronic structure and transport calculations.

We note that using this approach does not change our conclusions. Although the polarization of the fully relaxed structure is strongly suppressed, the structural asymmetry is still present in the system. This fact can be seen from comparison of the polar displacement for the fully relaxed structure (Fig. 8(b), top panel) to the heterostructure relaxed from the initial centrosymmetric configuration (Fig. 8(c), top panel). As a result, contrary to the centrosymmetric structure, we observe a non-zero ICHE in the fully relaxed structure (Figs. 8(b), bottom panel) comparable in magnitude to that for the heterostructure with the fixed bulk GeTe structure (Figs. 8(a), bottom panel). Below, we will focus on the properties of the GeTe/Ru$_2$MnGe/GeTe heterostructure relaxed with the fixed central Ge and Te atoms shown in Fig. 8(a).

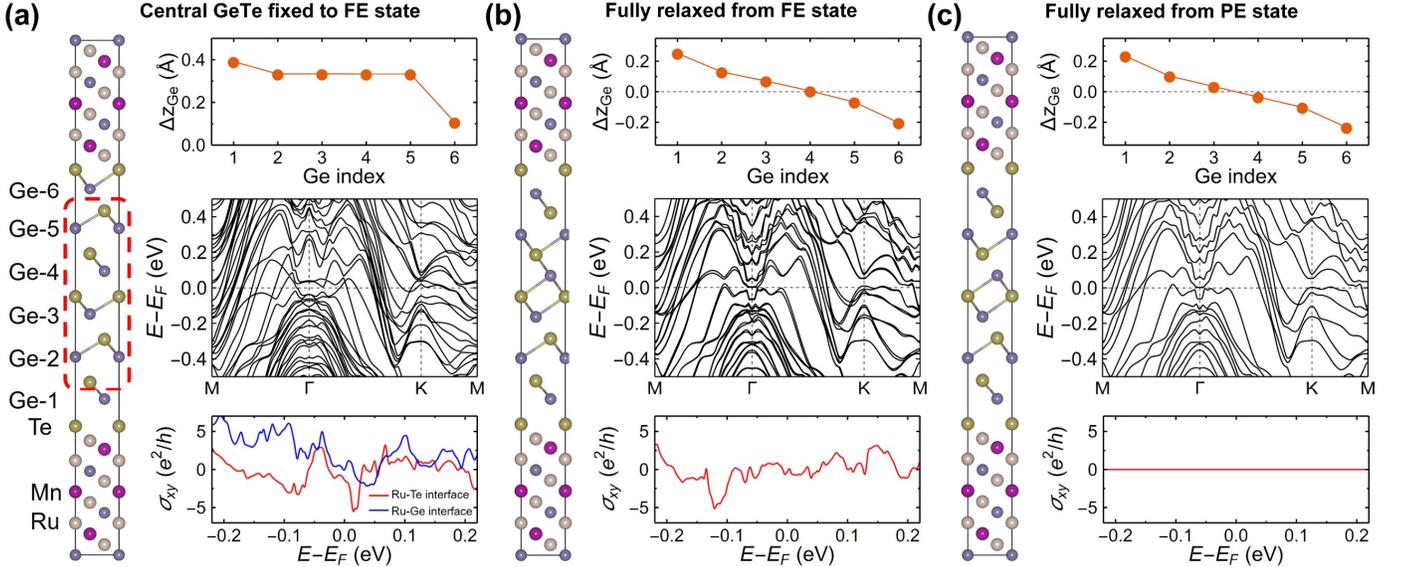

**FIG. 8:** (**a,b,c**) The GeTe/$Ru_2MnGe$/GeTe heterostructure (left panel), the polar displacements $\Delta z_{Ge}$ of the heterostructure (top of right panel), The calculated band structures (middle of right panel), and the calculated interfacial crystal Hall conductance (bottom of right panel) for the heterostructure relaxed with the central GeTe slab fixed to the bulk-like ferroelectric state (**a**), the heterostructure fully relaxed from the structure shown in (**a**) without constraint (**b**), and the heterostructure relaxed from with a centrosymmetric initial structure (**c**).

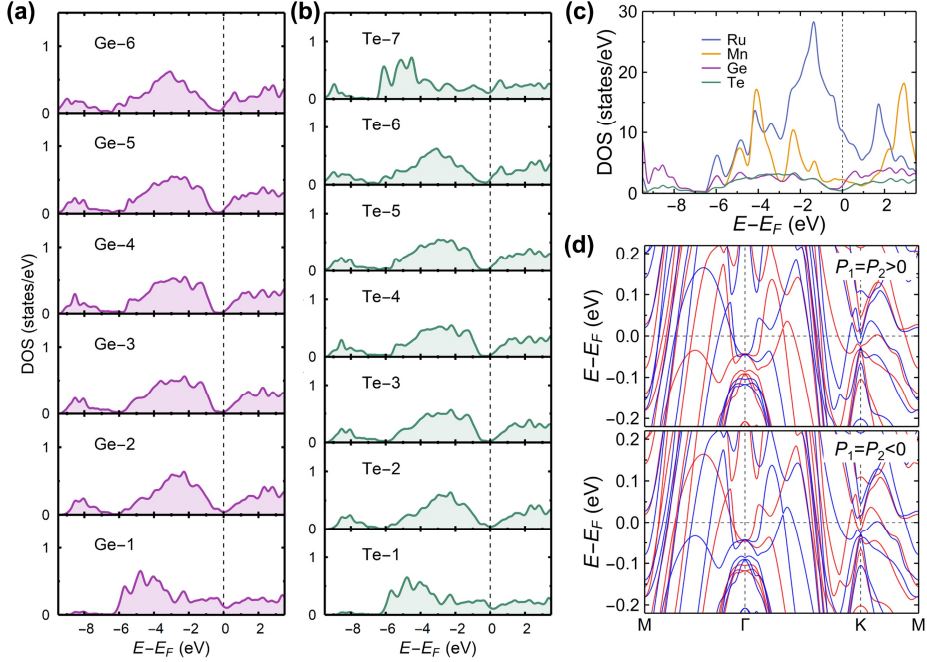

**FIG. 9:** Electronic structure of GeTe/Ru$_2$MnGe/GeTe. **(a,b)** The projected density of states (DOS) to Ge **(a)** and Te **(b)** layers. **(c)** The atom projected DOS. **(d)** The spin-dependent band structure polarizations $P_1$ and $P_2$ pointing up ($P_1 = P_2 > 0$, top panel) and down ($P_1 = P_2 < 0$, bottom panel) calculated without spin-orbit coupling. The red and blue lines denote the bands with up and down spin, respectively.

Figures 9(a-c) show the calculated layer (Figs. 9(a,b)) and atom (Fig. 9 (c)) projected density of states (DOS) in the GeTe/Ru$_2$MnGe/GeTe heterostructure. It is seen that the bulk-like band gap in the interior of the GeTe layer is well maintained (Figs. 9(a,b)). The DOS near the Fermi energy ($E_F$) is majorly contributed by Ru atoms and the contributions from Ge and Te atoms are very small (Fig. 9(c)). These facts indicate the ICHE largely arises from the electronic structure of Ru$_2$MnGe.

The heterostructure relaxed from the initial centrosymmetric structure is used to simulate the system with the top and bottom GeTe layers having opposite polarizations. In this case, the Kramer degeneracy leads to zero Berry curvature everywhere and enforces the vanishing interfacial crystal Hall effect (Fig. 8(c)).

## APPENDIX E: OTHER HETEROSTRUCTURES FOR REVERSIBLE ICHE

Below we list magnetic point group operations which enforce the antisymmetric relationship for the Berry curvature $\mathbf{\Omega}$ in 2D systems defined by Eq. (1):

1) $\hat{P}\hat{T}$, combination of space inversion and time reversal:

$$\hat{P}\hat{T}\mathbf{\Omega}(\boldsymbol{k}) = -\mathbf{\Omega}(\boldsymbol{k});$$

2) $\hat{M}_\parallel$, mirror reflection perpendicular to the $x$-$y$ plane:

$$\hat{M}_\parallel \mathbf{\Omega}(k_x, k_y) = -\mathbf{\Omega}(k'_x, k'_y);$$

3) $\hat{C}_{2\parallel}$, two-fold rotation around an in-plane direction:

$$\hat{C}_{2\parallel}\mathbf{\Omega}(k_x, k_y) = -\mathbf{\Omega}(k'_x, k'_y);$$

4) $\hat{T}\hat{C}_{2z}$, combination of time reversal and two-fold rotation around the $z$ axis:

$$T\hat{C}_{2z}\mathbf{\Omega}(\boldsymbol{k}) = -\mathbf{\Omega}(\boldsymbol{k});$$

5) $\hat{T}\hat{M}_z$, combination of time reversal and mirror reflection perpendicular to the $z$ direction:

$$\hat{T}\hat{M}_z\mathbf{\Omega}(\boldsymbol{k}) = -\mathbf{\Omega}(-\boldsymbol{k}).$$

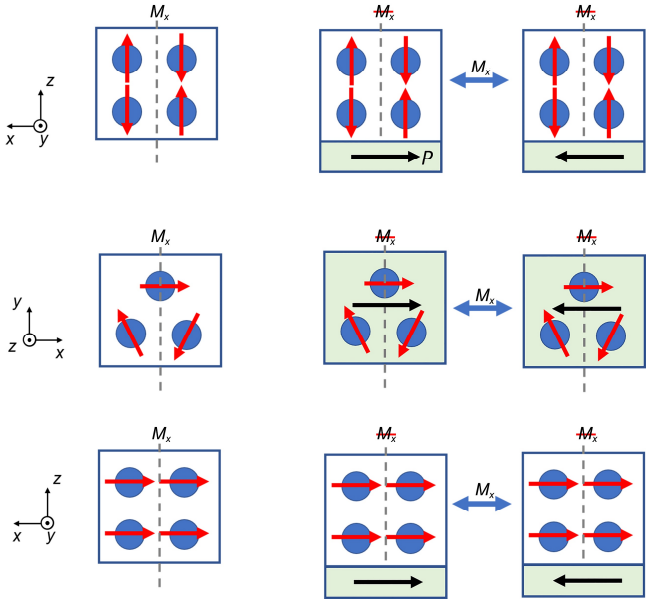

The design principles for the heterostructures to exhibit a ferroelectric-reversible ICHE are: (1) choose an AFM layer which has at least one of the symmetry operations listed above; (2) choose a ferroelectric layer, where the polarization breaks these symmetry operations and the polarization switching in this heterostructure is equivalent to the symmetry operation which changes sign of the Berry curvature.

Some examples of possible magnet/ferroelectric bilayer heterostructures are shown in Fig. 10. Here polarization of the ferroelectric layer is along the in-plane direction, and the magnetic metal layer has an in-plane AFM order, a non-collinear AFM order, or even a ferromagnetic order. In these magnets, mirror reflection $\hat{M}_x$ enforces zero in-plane anomalous Hall conductance due to $\hat{M}_x \mathbf{\Omega}(k_x, k_y) = -\mathbf{\Omega}(-k_x, k_y)$. In the heterostructures, the in-plane polarization along the $x$ direction breaks $\hat{M}_x$ and hence supports a non-zero ICHE. Polarization switching is equivalent to applying $\hat{M}_x$ to this system, which reverses the interfacial crystal Hall conductance.

**FIG. 10:** Three examples of the magnet/ferroelectric bilayer heterostructures supporting the ferroelectric controlled ICHE.

* dfshao@unl.edu
† tsymbal@unl.edu

[1] E. Y. Tsymbal and I. Žutić, Eds., *Spintronics Handbook: Spin Transport and Magnetism*, 2nd edition (CRC press, 2019).

[2] E. Y. Tsymbal, Electric toggling of magnets. *Nat. Mater.* **11**, 12 (2012).

[3] M. Fiebig, T. Lottermoser, D. Meier, and M. Trassin, The evolution of multiferroics, *Nat. Rev. Mater.* **1**, 16046 (2016).

[4] N. A. Spaldin and R. Ramesh, Advances in magnetoelectric multiferroics, *Nat. Mater.* **18**, 203(2019).

[5] C. A. F. Vaz, J. Hoffman, C. H. Ahn, and R. Ramesh, Magnetoelectric Coupling Effects in Multiferroic Complex Oxide Composite Structures, *Adv. Mater.* **22**, 2900 (2010).

[6] J. D. Burton and E. Y. Tsymbal, Prediction of electrically induced magnetic reconstruction at the manganite/ferroelectric interface, *Phys. Rev. B* **80**, 174406 (2009).

[7] Y. W. Yin, J. D. Burton, Y. -M. Kim, A. Y. Borisevich, S. J. Pennycook, S. M. Yang, T. W. Noh, A. Gruverman, X. G. Li, E. Y. Tsymbal and Q. Li, Enhanced tunnelling electroresistance effect due to a ferroelectrically induced phase transition at a magnetic complex oxide interface, *Nat. Mater.* **12**, 397 (2013).

[8] C. A. F. Vaz, J. Hoffman, Y. Segal, J. W. Reiner, R. D. Grober, Z. Zhang, C. H. Ahn, and F. J. Walker, Origin of the magnetoelectric coupling effect in Pb(Zr$_{0.2}$Ti$_{0.8}$)O$_3$/La$_{0.8}$Sr$_{0.2}$MnO$_3$ multiferroic heterostructure, *Phys. Rev. Lett.* **104**, 127202 (2010).

[9] R. O. Cherifi, V. Ivanovskaya, L. C. Phillips, A. Zobelli, I. C. Infante, E. Jacquet, V. Garcia, S. Fusil, P. R. Briddon, N. Guiblin, A. Mougin, A. A. Ünal, F. Kronast, S. Valencia, B. Dkhil, A. Barthélémy, and M. Bibes, Electric-field Control of Magnetic Order Above Room Temperature, *Nat. Mater.* **13**, 345 (2014).

[10] M. Fechner, P. Zahn, S. Ostanin, M. Bibes, and I. Mertig, Switching Magnetization by 180° with an Electric Field, *Phys. Rev. Lett.* **108**, 197206 (2012).

[11] G. Radaelli, D. Petti, E. Plekhanov, I. Fina, P. Torelli, B. R. Salles, M. Cantoni, C. Rinaldi, D. Gutiérrez, G. Panaccione, M. Varela, S. Picozzi, J. Fontcuberta and R. Bertacco, Electric control of magnetism at the Fe/BaTiO$_3$ interface, *Nat. Commun.* **5**, 3404 (2014).

[12] D. F. Shao, G. Gurung, T. R. Paudel, and E. Y. Tsymbal, Electrically reversible magnetization at the antiperovskite/perovskite interface, *Phys. Rev. Mater.* **3**, 024405 (2019).

[13] P. Lazić, K. D. Belashchenko, I. Žutić, Effective gating and tunable magnetic proximity effects in two-dimensional heterostructures, *Phys. Rev. B* **93**, 241401 (2016).

[14] J. Xu, S. Singh, J. Katoch, G. Wu, T. Zhu, I. Žutić, R. K. Kawakami, Spin inversion in graphene spin valves by gate-tunable magnetic proximity effect at one-dimensional contacts. *Nat. Commun.* **9**, 2869 (2018).

[15] L. L. Tao, E. Y. Tsymbal. Perspectives of spin-textured ferroelectrics, *J. Phys. D: Appl. Phys.* **54**, 113001 (2020).

[16] D. Di Sante, P. Barone, R. Bertacco, and S. Picozzi, Electric Control of the Giant Rashba Effect in Bulk GeTe, *Adv. Mater.* **25**, 509 (2013).

[17] M. Liebmann, C. Rinaldi, D. Di Sante, J. Kellner, C. Pauly, R. N. Wang, J. E. Boschker, A. Giussani, S. Bertoli, M. Cantoni, L. Baldrati, M. Asa, I. Vobornik, G. Panaccione, D. Marchenko, J. Sanchez-Barriga, O. Rader, R. Calarco, S. Picozzi, R. Bertacco, and M. Morgenstern, *Adv. Mater.* **28**, 560 (2016).

[18] R. Karplus and J. M. Luttinger, Hall effect in ferromagnetics, *Phys. Rev.* **95**, 1154 (1954).

[19] N. Nagaosa, J. Sinova, S. Onoda, A. H. MacDonald, and N. P. Ong, Anomalous Hall effect, *Rev. Mod. Phys.* **82**, 1539 (2010).

[20] M. V. Berry, Quantal phase factors accompanying adiabatic changes, *Proc. R. Soc. A* **392**, 45 (1984).

[21] M.-C. Chang and Q. Niu, Berry phase, hyperorbits, and the Hofstadter spectrum: Semiclassical dynamics in magnetic Bloch bands, *Phys. Rev. B* **53**, 7010 (1996).

[22] D. Xiao, M.-C. Chang, and Q. Niu, Berry phase effects on electronic properties, *Rev. Mod. Phys.* **82**, 1959 (2010).

[23] L. Šmejkal, R. González-Hernández, T. Jungwirth, and J. Sinova, Crystal Hall effect in collinear antiferromagnets, *Sci. Adv.* **6**, eaaz8809 (2020).

[24] Z. Feng, X. Zhou, L. Šmejkal, L. Wu, Z. Zhu, H. Guo, R. González-Hernández, X. Wang, H. Yan, P. Qin, X. Zhang, H. Wu, H. Chen, C. Jiang, M. Coey, J. Sinova, T. Jungwirth, and Z. Liu, Observation of the Crystal Hall Effect in a Collinear Antiferromagnet, *arXiv*:2002.08712 (2020).

[25] X. Li, A. H. MacDonald, and H. Chen, Quantum anomalous Hall effect through canted antiferromagnetism, *arXiv*:1902.10650 (2019).

[26] N. J. Ghimire, A. S. Botana, J. S. Jiang, J. Zhang, Y.-S. Chen, and J. F. Mitchell, Large anomalous Hall effect in the chiral-lattice antiferromagnet CoNb$_3$S$_6$, *Nat. Commun.* **9**, 3280 (2018).

[27] K. Samanta, M. Ležaić, M. Merte, F. Freimuth, S. Blügel, and Y. Mokrousov, Crystal Hall and crystal magneto-optical effect in thin films of SrRuO$_3$, *J. Appl. Phys.* **127**, 213904 (2020).

[28] P. Wadley, B. Howells, J. Železný, C. Andrews, V. Hills, R. P. Campion, V. Novák, K. Olejník, F. Maccherozzi, S. S. Dhesi, S. Y. Martin, T. Wagner, J. Wunderlich, F. Freimuth, Y. Mokrousov, J. Kuneš, J. S. Chauhan, M. J. Grzybowski, A. W. Rushforth, K. W. Edmonds, B. L. Gallagher, and T. Jungwirth, Electrical switching of an antiferromagnet, *Science* **351**, 587 (2016).

[29] M. Gotoh, M. Ohashi, T. Kanomata, and Y. Yamaguchi, Spin reorientation in the new Heusler alloys Ru$_2$MnSb and Ru$_2$MnGe, *Physica B* **213-214**, 306 (1995).

[30] T. Kanomata, M. Kikuchi, and H. Yamauchi, Magnetic properties of Heusler alloys Ru$_2$MnZ (Z = Si, Ge, Sn and Sb), *J. Alloys. Compd.* **414**, 1 (2006).

[31] G. Kádár, E. Krén, and M. Márton, New antiferromagnetic intermetallic compound in the Mn-Pd system: MnPd$_2$, *J. Phys. Chem. Solids.* **33**, 212 (1972).

[32] D.-F. Shao, G. Gurung, S.-H. Zhang, and E. Y. Tsymbal, Dirac nodal line metal for topological antiferromagnetic spintronics, *Phys. Rev. Lett.* **122**, 077203 (2019).

[33] D. G. Quirinale, V. K. Anand, M. G. Kim, Abhishek Pandey, A. Huq, P. W. Stephens, T. W. Heitmann, A. Kreyssig, R. J. McQueeney, D. C. Johnston, and A. I. Goldman, Crystal and magnetic structure of CaCo$_{1.86}$As$_2$ studied by x-ray and neutron diffraction, *Phys. Rev. B* **88**, 174420 (2013).

[34] M. M. Otrokov, I. I. Klimovskikh, H. Bentmann, D. Estyunin, A. Zeugner, Z. S. Aliev, S. Gaß, A. U. B. Wolter, A. V. Koroleva, A. M. Shikin, M. Blanco-Rey, M. Hoffmann, I. P. Rusinov, A. Yu. Vyazovskaya, S. V. Eremeev, Y. M. Koroteev, V. M. Kuznetsov, F. Freyse, J. Sánchez-Barriga, I. R. Amiraslanov, M. B. Babanly, N. T. Mamedov, N. A. Abdullayev, V. N. Zverev, A. Alfonsov, V. Kataev, B. Büchner, E. F. Schwier, S. Kumar, A.

Kimura, L. Petaccia, G. Di Santo, R. C. Vidal, S. Schatz, K. Kißner, M. Ünzelmann, C. H. Min, S. Moser, T. R. F. Peixoto, F. Reinert, A. Ernst, P. M. Echenique, A. Isaeva, and E. V. Chulkov, Prediction and observation of an antiferromagnetic topological insulator, *Nature* **576**, 416 (2019).

[35] B. Huang, G. Clark, E. Navarro-Moratalla, D. R. Klein, R. Cheng, K. L. Seyler, D. Zhong, E. Schmidgall, M. A. McGuire, D. H. Cobden, W. Yao, D. Xiao, P. Jarillo-Herrero and X. Xu, Layer-dependent ferromagnetism in a van der Waals crystal down to the monolayer limit, *Nature* **546**, 270 (2017).

[36] A. S. Avilov and R. M. Imamov, Electron-diffraction study of germanium diiodide, *Sov. Phys. Crystallogr.* **13**, 52 (1968).

[37] E. Urgiles, P. Melo, and C. C. Coleman, Vapor reaction growth of single crystal GeI$_2$, *J. Cryst. Growth* **165**, 245 (1996).

[38] C.-S. Liu, X.-L. Yang, J. Liu, and X.-J. Ye, Exfoliated monolayer GeI$_2$: Theoretical prediction of a wide-band gap semiconductor with tunable half-metallic ferromagnetism, *J. Phys. Chem. C* **122**, 22137 (2018).

[39] I. Sodemann and L. Fu, Quantum nonlinear Hall effect induced by Berry curvature dipole in time-reversal invariant materials, *Phys. Rev. Lett.* **115**, 216806 (2015).

[40] D.-F. Shao, S.-H. Zhang, G. Gurung, W. Yang, and E. Y. Tsymbal, Nonlinear anomalous Hall effect for Néel vector detection, *Phys. Rev. Lett.* **124**, 067203 (2020).

[41] C. Liu, Y. Wang, H. Li, Y. Wu, Y. Li, J. Li, K. He, Y. Xu, J. Zhang, and Y. Wang, Robust axion insulator and Chern insulator phases in a two-dimensional antiferromagnetic topological insulator, *Nat. Mater.* **19**, 522 (2020).

[42] Y. Deng, Y. Yu, M. Z. Shi, Z. Guo, Z. Xu, J. Wang, X. H. Chen, and Y. Zhang, Quantum anomalous Hall effect in intrinsic magnetic topological insulator MnBi$_2$Te$_4$, *Science* **367**, 895 (2020).

[43] D. Zhang, M. Shi, T. Zhu, D. Xing, H. Zhang, and J. Wang, Topological axion states in the magnetic insulator MnBi$_2$Te$_4$ with the quantized magnetoelectric effect, *Phys. Rev. Lett.* **122**, 206401 (2019).

[44] M. M. Otrokov, I. P. Rusinov, M. Blanco-Rey, M. Hoffmann, A. Y. Vyazovskaya, S. V. Eremeev, A. Ernst, P. M. Echenique, A. Arnau, and E. V. Chulkov, Unique thickness-dependent properties of the van der Waals interlayer antiferromagnet MnBi$_2$Te$_4$ films, *Phys. Rev. Lett.* **122**, 107202 (2019).

[45] J. Li, Y. Li, S. Du, Z. Wang, B. -L. Gu, S. -C. Zhang, K. He, W. Duan, and Y. Xu, Intrinsic magnetic topological insulators in van der Waals layered MnBi$_2$Te$_4$-family materials, *Sci. Adv.* **5**, eaaw5685 (2019).

[46] H. Fu, C.-X. Liu, and B. Yan, Exchange bias and quantum anomalous Hall effect in the MnBi$_2$Te$_4$/CrI$_3$ heterostructure, *Sci. Adv.* **6**, eaaz0948 (2020).

[47] S. Du, P. Tang, J. Li, Z. Lin, Y. Xu, W. Duan, and A. Rubio, Berry curvature engineering by gating two-dimensional antiferromagnets, *Phys. Rev. Research* **2**, 022025(R) (2020).

[48] W. Ding, J. Zhu, Z. Wang, Y. Gao, D. Xiao, Y. Gu, Z. Zhang, and W. Zhu, Prediction of intrinsic two-dimensional ferroelectrics in In$_2$Se$_3$ and other III$_2$-VI$_3$ van der Waals materials, *Nat. Commun.* **8**, 14956 (2017).

[49] J. Xiao, H. Zhu, Y. Wang, W. Feng, Y. Hu, A. Dasgupta, Y. Han, Y. Wang, D. A. Muller, L. W. Martin, P. A. Hu, and X. Zhang,

Intrinsic two-dimensional ferroelectricity with dipole locking, *Phys. Rev. Lett.* **120**, 227601 (2018).

[50] C. Cui, W.-J. Hu, X. Yan, C. Addiego, W. Gao, Y. Wang, Z. Wang, L. Li, Y. Cheng, P. Li, X. Zhang, H. N. Alshareef, T. Wu, W. Zhu, X. Pan, and L.-J. Li, Intercorrelated in-plane and out-of-plane ferroelectricity in ultrathin two-dimensional layered semiconductor In$_2$Se$_3$, *Nano Lett.* **18**, 1253-1258 (2018).

[51] P. B. Pereira, I. Sergueev, S. Gorsse, J. Dadda, E. Müller, and R. P. Hermann, Lattice dynamics and structure of GeTe, SnTe and PbTe, *Phys. Stat. Sol. B* **250**, 1300 (2013).

[52] Y. Deng, Y. Yu, Y. Song, J. Zhang, N. Z. Wang, Z. Sun, Y. Yi, Y. Z. Wu, S. Wu, J. Zhu, J. Wang, X. H. Chen, and Y. Zhang, Gate-tunable room-temperature ferromagnetism in two-dimensional Fe$_3$GeTe$_2$, *Nature* **563**, 94 (2018).

[53] X. Lin and J. Ni, Layer-dependent intrinsic anomalous Hall effect in Fe$_3$GeTe$_2$, *Phys. Rev. B* **100**, 085403 (2019).

[54] N. Fukatani, H. Fujita, T. Miyawaki, K. Ueda, and H. Asano, Structural and magnetic properties of antiferromagnetic Heusler Ru$_2$MnGe Epitaxial thin films, J. Kor. Phys. Soc. 63, 711 (2013).

[55] P. Blöchl, Projector augmented-wave method, *Phys. Rev. B* **50**, 17953 (1994).

[56] G. Kresse and D. Joubert, From ultrasoft pseudopotentials to the projector augmented-wave method, *Phys. Rev. B* **59**, 1758 (1999).

[57] J. P. Perdew, K. Burke, and M. Ernzerhof, Generalized gradient approximation made simple, *Phys. Rev. Lett.* **77**, 3865 (1996).

[58] V. I. Anisimov, J. Zaanen, and O. K. Andersen, Band theory and Mott insulators: Hubbard U instead of Stoner I, *Phys. Rev. B* **44**, 943 (1991).

[59] S. L. Dudarev, G. A. Botton, S. Y. Savrasov, C. J. Humphreys, and A. P. Sutton, Electron-energy-loss spectra and the structural stability of nickel oxide: An LSDA+U study, *Phys. Rev. B* **57**, 1505 (1998).

[60] S. Grimme, J. Antony, S. Ehrlich, and H. Krieg, A consistent and accurate ab initio parametrization of density functional dispersion correction (DFT-D) for the 94 elements H-Pu, *J. Chem. Phys.* **132**, 154104 (2010).

[61] A. D. Becke, E. R. Johnson, A simple effective potential for exchange, *J. Chem. Phys.* **124**, 221101 (2006).

[62] N. Marzari, A. A. Mostofi, J. R. Yates, I. Souza, and D. Vanderbilt, Maximally localized Wannier functions: Theory and applications, *Rev. Mod. Phys.* **84**, 1419 (2012).

[63] A. A. Mostofi, J. R. Yates, G. Pizzi, Y. S. Lee, I. Souza, D. Vanderbilt, and N. Marzari, An updated version of Wannier90: A tool for obtaining maximally-localised Wannier functions, *Comput. Phys. Commun.* **185**, 2309 (2014)

[64] Q. S. Wu, S. N. Zhang, H.-F. Song, M. Troyer, and A. A. Soluyanov, WannierTools: An open-source software package for novel topological materials, *Comput. Phys. Commun.* **224**, 405 (2018).

[65] J. R. Yates, X. Wang, D. Vanderbilt, and I. Souza, Spectral and Fermi surface properties from Wannier interpolation, *Phys. Rev. B* **75**, 195121 (2007).

[66] T. Williams and C. Kelley, Gnuplot 5.2: An interactive plotting program (2017), http://www.gnuplot.Info

[67] M. A. Caprio, LevelScheme: A level scheme drawing and scientific figure preparation system for Mathematica. *Comp. Phys. Commun.* **171**, 107 (2005).

[68] L. Bengtsson, Dipole correction for surface supercell calculations, *Phys. Rev. B* **59**, 12301-12304 (1999).

[69] F. Xue, Z. Wang, Y. Hou, L. Gu, and R. Wu, Control of magnetic properties of MnBi$_2$Te$_4$ using a van der Waals ferroelectric III$_2$-VI$_3$ film and biaxial strain, *Phys. Rev. B* **101**, 184426 (2020).

[70] J. Sławińska, D. Di Sante, S. Varotto, C. Rinaldi, R. Bertacco, and S. Picozzi, Fe/GeTe(111) heterostructures as an avenue towards spintronics based on ferroelectric Rashba semiconductors, *Phys. Rev. B* **99**, 075306 (2019).

[71] V. L. Deringer, M. Lumeij, and R. Dronskowski, Ab initio modeling of α-GeTe(111) surfaces, *J. Phys. Chem. C* **116**, 15801 (2012).